\documentclass[a4paper,twoside]{article}

\usepackage{cite}
\usepackage{url}
\usepackage{epsfig}
\usepackage{subcaption}
\usepackage{calc}
\usepackage{amssymb}
\usepackage{amstext}
\usepackage{amsmath}
\usepackage{amsthm}
\usepackage{multicol}
\usepackage{pslatex}
\usepackage{apalike}
\usepackage{algorithm2e}
\usepackage[bottom]{footmisc}

\usepackage[]{mdframed}

\newcommand{\ra}{\rangle}

\newcommand{\compose}{\parallel}

\newcommand{\request}{{\color{blue}request}}
\newcommand{\waitfor}{{\color{green!50!black}wait for}}
\newcommand{\blocking}{{\color{red}blocking}}

\usepackage{tikz}
\usetikzlibrary{arrows,trees,backgrounds,automata,shapes,decorations,plotmarks,fit,calc,positioning,shadows,chains,
  patterns,decorations.pathreplacing}
\tikzstyle{block}=[rectangle, draw, thin, inner sep=3pt, text centered, drop shadow, fill=orange!20!yellow!20]
\tikzstyle{pre}=[<-,shorten <=1pt,>=stealth']
\tikzstyle{post}=[->,shorten >=1pt,>=stealth']
\tikzstyle{bi}=[<->,shorten >=1pt,shorten <=1pt,>=stealth']
\tikzstyle{every initial by arrow}=[initial text={},initial distance=1em,post]
\tikzstyle{every state}=[minimum size=0.4cm,drop shadow,fill=orange!20!yellow!20]
\tikzstyle{transition}= [post,shorten >=1pt,node distance=2cm, inner sep=2pt,bend angle=20]
\usetikzlibrary{calc}

\usepackage{tikz,ifthen,pgfplots}
\pgfplotsset{compat=1.18}
\usetikzlibrary{arrows,trees,backgrounds,automata,shapes,decorations,plotmarks,fit,calc,positioning,shadows,chains}
\tikzstyle{every pin edge}=[<-,shorten <=1pt]
\tikzstyle{neuron}=[circle,fill=black!25,minimum size=17pt,inner sep=0pt]
\tikzstyle{input neuron}=[neuron, fill=green!50]
\tikzstyle{output neuron}=[neuron, fill=red!50]
\tikzstyle{hidden neuron}=[neuron, fill=blue!50]
\tikzstyle{annot} = [text width=6em, text centered]
\usetikzlibrary{calc}

\DeclareFixedFont{\ttb}{T1}{txtt}{bx}{n}{7.5} 
\DeclareFixedFont{\ttm}{T1}{txtt}{m}{n}{7.5}  

\definecolor{deepblue}{rgb}{0,0,0.5}
\definecolor{deepred}{rgb}{0.6,0,0}
\definecolor{deepgreen}{rgb}{0,0.5,0}

\usepackage{SCITEPRESS}     

\begin{document}
\title{On Augmenting Scenario-Based Modeling with Generative AI}

\author{\authorname{David Harel\sup{1}, Guy Katz\sup{2}, Assaf
    Marron\sup{1}, and Smadar Szekely\sup{1}}
  \affiliation{\sup{1}Weizmann Institute of Science, Rehovot, Israel}
  \affiliation{\sup{2}The Hebrew University of Jerusalem, Jerusalem, Israel}
\email{dharel@weizmann.ac.il, guykatz@cs.huji.ac.il,
  assaf.marron@weizmann.ac.il, smadarsz@gmail.com}
}

\keywords{Generative AI, chatbots, scenario-based modeling, rule-based
  specifications}

\abstract{The manual modeling of complex systems is a daunting task; and although a plethora of methods exist that mitigate this issue, the problem remains very difficult. Recent advances in generative AI have allowed the creation of general-purpose chatbots, capable of assisting software engineers in various modeling tasks. However, these chatbots are often inaccurate, and an unstructured use thereof could result in erroneous system models. In this paper, we outline a method for the safer and more structured use of chatbots as part of the modeling process. To streamline this integration, we propose leveraging scenario-based modeling techniques, which are known to facilitate the automated analysis of models. We argue that through iterative invocations of the chatbot and the manual and automatic inspection of the resulting models, a more accurate system model can eventually be obtained. We describe favorable preliminary results, which highlight the potential of this approach.}

\onecolumn \maketitle \normalsize \setcounter{footnote}{0} \vfill

\section{\uppercase{Introduction}}
\label{sec:introduction}

Manually modeling complex systems is a daunting and error-prone
endeavor. Furthermore, even after the system is modeled, ongoing tasks,
such as modification and repair, continue to tax human
engineers. Creating tools and methodologies for streamlining and
facilitating this process has been the topic of extensive work, but
many aspects of the problem remain unsolved~\cite{PeAn16,BiMiNaTr05}.

In recent years, the deep learning revolution has been causing
dramatic changes in many areas, including computer science; and this revolution has recently
taken yet another step towards general-purpose AI, with the release of
ChatGPT, the learning-based chatbot~\cite{chatgpt}. ChatGPT, and
other, similar tools~\cite{bard,llama}, can be used for countless
kinds of tasks --- including the modeling and coding of complex
systems~\cite{SuSh23}. An engineer might provide ChatGPT with a
natural-language description of the system at hand, and receive in
return a model of the system, or even computer code that implements
it; and through iterative querying of ChatGPT, the system can later be
modified or enhanced. This approach has already been used in
several application domains~\cite{SuSh23,BuOzAyTu23,LiXiWaZh23}.

Although the ability to integrate ChatGPT\footnote{We will often use the term \emph{ChatGPT} somewhat generically, to represent an arbitrary, modern chatbot.} into the software development cycle will undoubtedly
empower engineers, there are also potential pitfalls that need to be
taken into account.  One drawback of ChatGPT and similar tools is that
the answers they provide are often inaccurate, and might overlook
important aspects of the input query~\cite{LiXiWaZh23}. Moreover, the
input query itself might be imperfect, and the engineer might not
realize this until the system is deployed.  Thus, if we make the
reasonable assumption that human engineers will gradually become
dependent on chatbots for various tasks, the risk increases that
these inaccuracies will find their way into the final models of the
system at hand and the code that ensues. 
We are thus faced with the following challenge: how can we harness
ChatGPT in a way that  lifts a significant load of work off the
shoulders of the engineers, but which  still results in sound and
accurate models?

Here, we advocate the creation of an encompassing modeling scheme that
will combine ChatGPT with more traditional techniques for manual
modeling of systems~\cite{BiMiNaTr05,PeAn16}, in a way that will
achieve this goal.  Our core idea is to use ChatGPT in a controlled
way; i.e., to repeatedly invoke it for various tasks, but to then
thoroughly inspect and analyze its results, to ensure their soundness
and accuracy. We argue that such a scheme, if designed properly, would
allow software and system engineers to benefit from the capabilities
of modern chatbots, but without jeopardizing the quality of the
resulting products.
In the long run, we regard such a scheme as
a step towards the \emph{Wise Computing}
vision~\cite{HaKaMaMa18}, which calls for turning the computer into a
proactive member of the software development team --- one which can
propose courses of action, detect under-specified portions of the
model, and assist in the various routine actions that naturally arise
as part of the software development cycle.

In order to design such a modeling scheme, we propose to leverage the
extensive work carried out in the modeling community over the years.
Specifically, we propose to focus on modeling frameworks that afford
two benefits that complement the capabilities of ChatGPT: (i) the
models produced by the framework are naturally well-aligned with how
humans perceive systems; this, we believe, will make it easier for the
human engineer to inspect ChatGPT's output; and (ii) the
resulting models are amenable to automated analysis tasks, such as
model checking, which will support the automated detection of bugs and
inconsistencies in the automatically generated models.  

Several modeling approaches fit this description, and many of them can probably
be used, but for the initial evaluation presented here, we focus on
\emph{scenario-based modeling} (\emph{SBM}) --- a technique that
generates models comprised of simple \emph{scenarios}, each of which
describes a single aspect of the system at
hand~\cite{HaMaWe12,DaHa01}. As we later discuss, this can
facilitate the smooth collaboration between ChatGPT and the human engineers.

To demonstrate the potential of this combined framework, we focus on
a few tasks that  arise naturally as part of a system's life
cycle. Specifically, we discuss the initial design of the model, its testing and the 
verification of its properties, 
its later enhancement
or repair due to the discovery of inconsistencies, and also a search for under-specified portions of the
model. Our results, although preliminary, are very promising, and we
hope this paper will form a basis for further research in this
direction.



In the remainder of the paper, we present the key concepts of our
approach, and discuss a high-level plan for the next steps. We begin by
introducing the concepts of SBM and language model-based chatbots in
Section~\ref{sec:background}. Next, we present the proposed integration of
SBM and ChatGPT in Section~\ref{sec:integration}, followed by a discussion of
some of the more advanced aspects of this integration in
Section~\ref{sec:advancedIntegration}. We discuss related work in
Section~\ref{sec:relatedWork} and conclude in Section~\ref{sec:conclusion}.

\section{\uppercase{Background}}
\label{sec:background}

\subsection{Large Language Model-Based Chatbots}
ChatGPT (Chat Generative Pre-trained Transformer) is a large language
model (LLM) based chatbot, developed by
OpenAI~\cite{chatgpt,ChWaWaWuZhChYaYiWaWaYe23}. The chatbot is able to
conduct an iterative conversation of variable length, format, style,
level of detail, and language. At each stage, the user presents a new
prompt to ChatGPT, which then replies, based on all previous prompts in
that conversation (the context). Following its debut in 2022, ChatGPT
quickly became highly successful, and inspired multiple other companies to
release their own chatbots~\cite{bard,llama}.

Internally, ChatGPT is implemented using a proprietary series of
generative pre-trained transformer (GPT) models, which in turn are
based on Google's transformer
architecture~\cite{VaShPaUsJoGoKaPo17}. ChatGPT is fine-tuned for
conversational applications, through a combination of supervised and
reinforcement learning techniques, as well as manual adjustment by
human engineers.  ChatGPT's training, as well as its inference, are
considered very costly in terms of power consumption and processing
resources.

Functionality-wise, ChatGPT is highly versatile. Some of its many uses
include generating student
essays~\cite{AlDiJoLo23},  writing and debugging computer
programs~\cite{SuSh23}, and composing music~\cite{LuXuKaYuXiTaBi23}. However, it will sometimes produce
plausible-sounding but incorrect or nonsensical answers --- a common
limitation for large language models~\cite{GrPe23}.

\subsection{Scenario-Based Modeling}

Scenario-based modeling~\cite{HaMaWe12} (SBM) is a modeling approach
aimed at modeling complex, reactive systems. The main component in a
scenario-based (SB) model is the \emph{scenario object}, which
describes a single behavior of the system at hand, whether desirable
or undesirable, so that one can specify it as necessary, allowed or forbidden. 
Each scenario object does not directly interact with
its counterparts, and can be created in isolation. Cross-scenario
interaction is allowed only through a global execution mechanism,
which can execute a collection of scenarios in a manner that produces
cohesive, global behavior.

There exist several flavors of SBM, employing slightly
different mechanisms for cross-scenario interactions.
We focus here on a particular set of idioms,
which has become quite popular: the \emph{requesting}, \emph{waiting-for} and
\emph{blocking} of discrete events~\cite{HaMaWe12}. During execution,
each scenario object repeatedly visits designated
\emph{synchronization points}, and in each of these the global
execution mechanism  selects one event for triggering. A
scenario object may declare events that it wishes to be triggered
(\emph{requested} events), events that it wishes to avoid
(\emph{blocked} events), and also events it does not request itself
but would like to monitor (\emph{waited-for events}). The execution
mechanism collects these declarations from each of the scenario
objects (or a subset thereof~\cite{HaKaKa13}), selects one event that
is requested and not blocked, and then informs all relevant scenario
objects of this selection.

In a given synchronization point, multiple events 
may be requested and not blocked, and several strategies have been proposed for
selecting one of them. These include an arbitrary selection, a
random selection, a round-robin mechanism, and 
look-ahead that simulates possible progression of the execution and selects events with an attempt to 
achieve a desirable objective specified a-priori (e.g., the avoidance
of deadlocks).
Executing a scenario-based program in this manner is termed play-out~\cite{HaMa03}).

\begin{figure}[htp]
  \centering
  \scalebox{0.65} {
    
    \tikzstyle{box}=[draw,  text width=2cm,text centered,inner sep=3]
    \tikzstyle{set}=[text centered, text width = 10em]

    \begin{tikzpicture}[thick,auto,>=latex',line/.style ={draw, thick, -latex', shorten >=0pt}]
      
      \matrix(bts) [row sep=0.3cm,column sep=2cm]  {

        \node (box1)  [box] {\waitfor{} \textsc{WaterLow}}; \\
        \node (box2)  [box] {\request\ \textsc{AddHot}}; \\
        \node (box3)  [box] {\request\ \textsc{AddHot}}; \\ 
        \node (box4)  [box] {\request\ \textsc{AddHot}}; \\ 
      };

      \draw [->] ($(box1.north) + (0,0.3cm)$) -- (box1.north);
      \node (title) [above=0.1cm of bts,box,draw=none] at ($(bts) + (-0.25cm,2.51cm)$) 
      {\textsc{AddHotWater}};  
      
      \begin{scope}[every path/.style=line]
        \path (box1)   -- (box2);
        \path (box2)   -- (box3);
        \path (box3)   -- (box4);
        \path (box4.east)   -- +(.25,0) |- (box1);
      \end{scope}

      \matrix(bts2) [right=.25cm of bts, row sep=0.3cm,column sep=2cm] {
        \node (box1)  [box] {\waitfor{} \textsc{WaterLow}}; \\
        \node (box2)  [box] {\request\ \textsc{AddCold}}; \\
        \node (box3)  [box] {\request\ \textsc{AddCold}}; \\ 
        \node (box4)  [box] {\request\ \textsc{AddCold}}; \\ 
      };
      
      \draw [->] ($(box1.north) + (0,0.3cm)$) -- (box1.north);
      \node (title) [above=0.1cm of bts2,box,draw=none] at ($(bts2) + (-0.25cm,2.51cm)$) 
      {\textsc{AddColdWater}};

      \begin{scope}[every path/.style=line]
        \path (box1)   -- (box2);
        \path (box2)   -- (box3);
        \path (box3)   -- (box4);
        \path (box4.east)   -- +(.25,0) |- (box1);
      \end{scope}

      \matrix(bts3) [right=.25cm of bts2, row sep=0.3cm,column sep=2cm] {
        \node (box1)  [box] {\waitfor{}  \textsc{AddHot} while  \blocking\ \textsc{AddCold}}; \\
        \node (box2)  [box] {\waitfor{}  \textsc{AddCold} while \blocking\ \textsc{AddHot}}; \\
      };

      \draw [->] ($(box1.north) + (0,0.3cm)$) -- (box1.north);
      \node (title) at (title-|bts3) [box,draw=none] {\textsc{Stability}};  

      \begin{scope}[every path/.style=line]
        \path (box1)   -- (box2);
        \path (box2.east)   -- +(.25,0) |- (box1);
      \end{scope}
      
      \node (log)  [right=.3cm of bts3,box,text width=2cm,fill=yellow!20] {
        $\cdots$ \\ 
        \textsc{WaterLow} \\
        \textsc{AddHot}  \\ 
        \textsc{AddCold} \\ 
        \textsc{AddHot}  \\ 
        \textsc{AddCold} \\ 
        \textsc{AddHot}  \\ 
        \textsc{AddCold} \\ 
        $\cdots$ \\
      }; 

      \node (title2) at (title-|log)            
      [box,draw=none] {\textsc{Event Log}};  
    \end{tikzpicture}
  }  
  \caption{
        A scenario-based model for a
    system that controls the water level in a tank with hot and
    cold water taps (taken from~\cite{HaKaMaWe14}).
  }  
  \label{fig:watertap}
\end{figure}
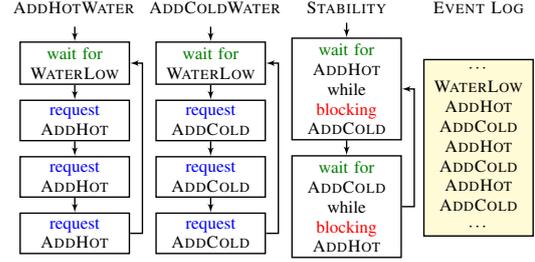

Fig.~\ref{fig:watertap} depicts a simple example of an SB model. The
system at hand controls the water level in a water tank, which is equipped with
hot and cold water taps. Each scenario object appears as a transition
system, in which nodes corresponds to the predetermined
synchronization points.  Scenario object \textsc{AddHotWater}
repeatedly waits for \textsc{WaterLow} events, and when such an event is
triggered, it requests three times the
event \textsc{AddHot}. Similarly, scenario object \textsc{AddColdWater} requests
the addition of cold water. When the model  includes
only objects \textsc{AddHotWater} and \textsc{AddColdWater}, three \textsc{AddHot}
events and three \textsc{AddCold} events may be triggered in any order
during execution.  If we wish to maintain a more stable water
temperature within the tank, we might add the scenario object
\textsc{Stability}, to enforce the interleaving of \textsc{AddHot} and
\textsc{AddCold} --- through the use of event blocking. An execution trace
of the model containing all three objects appears in the event log.

The SBM framework has been implemented on top of multiple high-level
languages, including Java~\cite{HaMaWe10}, C++~\cite{HaKa14},
Python~\cite{Ya23}, JavsScript~\cite{BaWeRe18} and
ScenarioTools~\cite{GeGrGuKoGlMaKa17}. Furthermore, SBM has been
applied in modeling various complex systems, such as
web-servers~\cite{HaKa14}, cache coherence protocols~\cite{HaKaMaMa16}
and robotic controllers~\cite{GrGr18b}. In order to simplify the
presentation in the following sections, we mostly describe SB models
as transitions systems.

In formally defining SBM, we follow the definitions of~\cite{Ka13}.
A scenario object
$O$ over event set $E$ is a tuple
$O = \langle Q, \delta, q_0, R, B \ra$, where the components are
interpreted as follows:
\begin{itemize}
\item $Q$ is the set of states. Each state represents a single,
  predetermined synchronization point;
\item $q_0\in Q$ is the initial state;
\item $R:Q\to 2^E$ and $B:Q\to 2^E$
map states to the sets of events requested and blocked at these
states, respectively; and
\item $\delta: Q \times E \to 2^Q$ is the transition function, which
  indicates how the object switches states in response to the
  triggering of events.
  \end{itemize}

Once the individual scenario objects are created, they can be composed
 in a pairwise fashion. 
Two scenario objects
 $O^1 = \langle Q^1, \delta^1, q_0^1, R^1, B^1 \ra$ and
 $O^2 = \langle Q^2, \delta^2, q_0^2, R^2, B^2 \ra$, specified over a
 common set of events $E$, can be composed into a single scenario object
$O^1 \compose O^2 = \langle Q^1 \times Q^2, \delta,
\langle q_0^1,q_0^2\rangle, R^1\cup R^2, B^1\cup B^2\rangle
$,
where:
\begin{itemize}
\item $\langle \tilde{q}^1,\tilde{q}^2\rangle \in  \delta(\langle q^1,q^2\rangle, e)$
if and only if $\tilde{q}^1 \in \delta^1(q^1,e)$ and $\tilde{q}^2\in
\delta^2(q^2,e)$; and
\item the union of the labeling functions is defined in the natural way; i.e., $e\in (R^1\cup
R^2)(\langle q^1,q^2 \rangle)$ if and only if $e \in R^1(q^1) \cup
R^2(q^2)$, and
$e\in (B^1\cup
B^2)(\langle q^1,q^2 \rangle)$ if and only if $e \in B^1(q^1) \cup
B^2(q^2)$.
\end{itemize}

Using the composition operator $\compose$, we can define a \emph{behavioral
  model} $M$ as a collection of scenario objects,
$M=\{O^1, O^2,\ldots, O^n\}$.  The executions of $M$ are defined to be
the executions of the single, composite object
$O = O^1\compose O^2\compose\ldots\compose O^n$.  Thus, each execution
starts from the initial state of $O$, which is the $n$-tuple of the initial states of 
its constituent objects, and throughout the run, in each
state $q$ an enabled event $e\in R(q)-B(q)$ is chosen for triggering,
if one exists.  The execution then moves to a state
$\tilde{q}\in \delta(q,e)$, and the process repeats.


\section{\uppercase{Integrating ChatGPT and SBM}}
\label{sec:integration}

\subsection{Basic Integration}

As a first step to integrating ChatGPT and SBM, we present a simple
methodology for creating scenario objects from free-text, using
ChatGPT. In order to get ChatGPT to present its output as a scenario
object, we propose to include in each query a \emph{preamble} that
describes the SBM semantics, as described in
Section~\ref{sec:background}. This preamble instructs ChatGPT to present
its output as a transition system, in which each state is labeled with
requested and blocked events, and has a set of transition rules
indicating its successor state, based on the event later triggered.
Note that this preamble needs only be crafted once, and can then be
attached to any subsequent query. If a specific
implementation of SBM is used (e.g., C++, Java), the output can be
crafted to match the required syntax of that implementation.

The simple preamble that we used in our initial
experiments is as follows:
\begin{mdframed}[backgroundcolor=blue!10]
I would like you to help me create a scenario-based model. A
scenario-based model is comprised of a finite set of events, and
scenarios. Each scenario is a sequence of states. In each state the
scenario can request events, wait-for events, and block events.
Afterwards, one event that is requested by at least one scenario, but
is blocked by no scenario, is selected for triggering. Scenarios that
requested or waited-for this scenario may then change states, and the
process repeats itself. Scenarios cannot have internal loops or
variables; these must be expressed using additional states.

To display scenarios, please use the following format. For each state,
mention the state's name, and the sets of requested and blocked
events. Also, mention the transitions to other states. For example:

\begin{itemize}
  \item
    s1: request X, block Y. If X is triggered, go to state s2.
  \item
    s2: request Y, block X. If Y is triggered, go to state s2.
\end{itemize}

Now, here are the details of the specific scenario-based model I'd
like you to create\ldots
\end{mdframed}
Given this preamble, we can use ChatGPT to produce the model we described in
Section~\ref{sec:background}. To produce the \textsc{AddHotWater} scenario, we
use the following prompt:

\begin{mdframed}[backgroundcolor=blue!10]
Consider the event set {WaterLow, AddHot, AddCold}.

Please suggest a scenario that waits for event WaterLow (without
requesting it). When WaterLow is triggered, the scenario performs the
following, 3 times in a row: request the event AddHot, and wait for it
to be triggered (both in a single state). When the 3rd AddHot event
occurs, the scenario transitions back to its initial state. 
\end{mdframed}
In response to this prompt, the chatbot replies:
\begin{mdframed}[backgroundcolor=purple!10]
  \begin{itemize}
    \item
      s1: wait for WaterLow. If WaterLow is triggered, go to state s2.
    \item 
s2: request AddHot, wait for AddHot. If AddHot is triggered, go to
state s3.
\item 
s3: request AddHot, wait for AddHot. If AddHot is triggered, go to
state s4.
\item 
s4: request AddHot, wait for AddHot. If AddHot is triggered, go to
state s1.
\end{itemize}
\end{mdframed}
Which is equivalent to the transition system we presented in
Fig.~\ref{fig:watertap}. The \textsc{AddColdWater} scenario can be generated
symmetrically. Now, if we wish to incrementally modify the system by
introducing the \textsc{Stability} scenario, we can issue the following prompt:
\begin{mdframed}[backgroundcolor=blue!10]
Please suggests a scenario that ensures that uses blocking to ensure
that no two consecutive AddHot events can be triggered, and that no
two consecutive AddCold events can be triggered; that is, once AddHot
is triggered, AddCold must be triggered before AddHot can be triggered
again, and vice versa. This scenario should not request any events,
and should work regardless of any WaterLow events. 
\end{mdframed}
And in response, the chatbot produces the
\textsc{Stability} scenario, as described in Fig.~\ref{fig:watertap}.

We note a subtle difference between the way we prompted ChatGPT for
the first two scenarios, \textsc{AddHotWater} and \textsc{AddColdWater}, and our
prompting for \textsc{Stability}. In the former two cases, our prompt
contained information that roughly described the transition system
itself, whereas in the third case our description was more
high-level, and did not mention the word ``state''. Still, in both cases,
ChatGPT produced the desired result. This demonstrates the
wide range of specifications with which the chatbot can successfully deal, and 
suggests that it can
be used even when the engineers are themselves not entirely certain of
the structure of the scenario they require. While it stands to reason
that more accurate descriptions would lead to more accurate results,
it appears that even high-level descriptions can be very useful, especially when combined
with the automated analysis techniques that we discuss next.

\subsection{The Proposed Methodology}

Building upon this basic integration of ChatGPT and SBM, we now
outline a structured LLM-agnostic and language-agnostic methodology
for creating complex \emph{reactive models}, i.e., models of systems
that interact with their environment repeatedly over time, and receive
external inputs~\cite{HaPn85}. Numerous modern,
critical systems can be regarded as reactive~\cite{AcInLaSr07}, and consequently there has been extensive research on 
developing tools and methods for modeling these systems. Despite this tremendous effort, there still remain significant gaps, which could result in models that are inaccurate or that are difficult to maintain. 
The present work, which can be seen as an element of the Wise Computing 
vision~\cite{HaKaMaMa18}, seeks to mitigate these gaps, through the creation of advanced, intelligent tools that will begin to undertake the software and system development tasks that are normally assigned to humans. 
The core of the approach is to have 
system components be generated,
iteratively and incrementally,
with the help of an LLM; and to have the LLM's
outputs checked systematically, and perhaps automatically, using
various tools and methods.
\begin{enumerate}
\item Describe the problem and the environment textually, in natural language.  
\item Choose a compositional, scenario-based modeling language, which
  has well-defined execution semantics and is suitable for
  the incremental development of the system.
\item Obtain an LLM that is familiar with the application domain in
  general, or can readily gain extensive knowledge about that domain,
  and which can (or can be taught to) generate code in the chosen
   scenario-based language.
\item Describe, perhaps iteratively, the semantics of the
  scenario-based language to the LLM as a preamble. Confirm that the LLM indeed
  internalizes the details of the language semantics by teaching it to
  execute (i.e., play out~\cite{HaMa03}) systems described as
  scenarios or rules in the chosen language, where the LLM outputs logs
  of triggered events, scenario states, composite system states,
  values of environment variables and changes thereto, etc.
\item Iteratively add scenarios and refine existing ones, as follows:
\begin{enumerate}	
\item Describe in a prompt one or more scenarios for certain not-yet-specified
  requirements or aspects of the system.
\item Have the LLM generate  actual scenarios for the prompt, in the
  chosen language.
\item Have the LLM generate natural language description of properties
  to be verified, executable test cases, and assertions for formal
  verification tools, per the original requirements. This constitutes
  stating the requirement at hand from different perspectives.
\item Carry out initial testing and validation within the LLM,
  challenging the LLM to find gaps and incorrect execution paths on its
  own. Correct the natural language specification and prompts as
  needed.
\item Systematically check the LLM output outside of the LLM, using
  any or all of the following: code reviews by human engineers, unit
  testing of individual scenarios, subsystem testing with some or all
  of the already-developed scenarios, model checking of the new
  scenarios, as well as those of the composite system, etc.  The testing is
  to be carried out in the execution environment of the language, and
  model checking is to be carried out using a suitable formal
  verification tool. Both should be independent of the LLM
  environment.  Possibly automate the subjecting of generated
  scenarios to testing and model checking.
\item When errors are found, do not modify the generated
  code. Instead, revise the LLM prompts until correct system scenarios
  and verification and testing properties are generated.  This step is critical
  for aligning the stakeholder (i.e., customer) view of the requirements,
  the developer's understanding, and the actual code.
\item Once the set of generated scenarios seems ready, repeat the likes of step (d),
  asking the LLM to find gaps or potential failures in this set of
  scenarios; specifically look for LLM suggestions of new environment
  considerations that prevent the system from working correctly. This
  step simulates the common system engineering task of having external
  experts or potential customers review advanced prototypes of
  systems. Repeat earlier steps as needed.
\end{enumerate}
\end{enumerate}

Next, we elaborate on some of these steps, and provide
simple, illustrative examples.

\section{\uppercase{Using the Method in the Development Cycle}}
\label{sec:advancedIntegration}

\subsection{Code Generation}

Code generation is probably the most straightforward chatbot capability that we propose be integrated into the development cycle. In
Section~\ref{sec:integration} we showed that ChatGPT can generate an
(executable) SB model --- a capability that has also been demonstrated with
other languages~\cite{SuSh23,BuOzAyTu23,LiXiWaZh23}.  A unique
advantage in the context of SB systems is the ability to generate
stand-alone scenarios, which can be reviewed and tested separately,
and then be incrementally added to the system under development. In
our preliminary testing for this paper, we experimented with code
generation for requirements in the realms of autonomous vehicles,
algorithms on data structures, simulating natural phenomena, and
control systems. In all of these, the ChatGPT/SBM integration proved
useful.

\subsection{Modeling}

Once ChatGPT understood the principles underlying scenario-based models, it
was able to combine its knowledge of the problem domain, the world at
large, and logic, in order to develop or enhance a model. It was able
to introduce new environment events, describe the sensor scenarios
that are required for triggering these events, and then add the
necessary application scenarios that react to these events. For
example, when we asked ChatGPT to generate scenario-based code for a
bubble-sort algorithm to be used by a robot moving numbered tiles on a sequence
of cells, it was able to introduce the events of detecting the arrival of a tile at
the tail-end of the array, as well as scenarios for reacting to such events.

\subsection{Play Out \& Simulation}
\begin{figure*}[ht]
    \centering
    \includegraphics[scale=0.7]{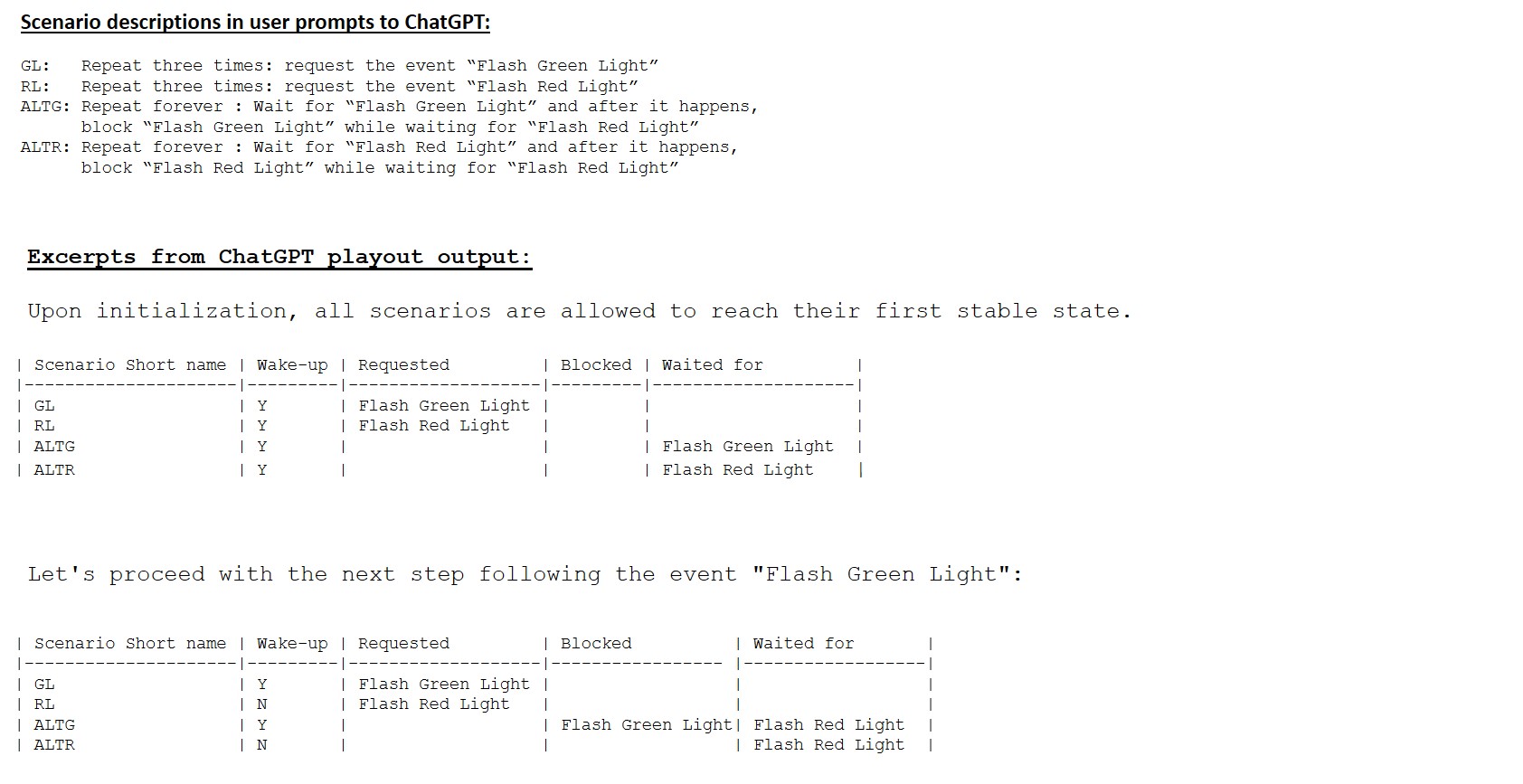}
    \caption{\emph{A simulation of scenario execution with ChatGPT.} Top: textual description of scenarios in user prompts. Middle and bottom: The first two entries in ChatGPT's output when requested to play out the execution, showing triggered events, the scenarios that reacted to the triggered events, and the declarations of each scenario at the next synchronization point. }
    \label{fig:playout}
\end{figure*}

After a few iterations, we were able to teach ChatGPT to produce an
execution log of an arbitrary scenario-based specification. At first
we observed ``wishful thinking'', where ChatGPT described the run as
it should be per the problem description.
However, as illustrated in Fig.~\ref{fig:playout},
ChatGPT was then able to follow the execution steps correctly,
displaying at each synchronization point the event that triggered the state transition that led to this synchronization point, and a table of all scenarios,
indicating for each one whether or not it was put into action by the triggered event,
and providing its declaration of requested, blocked and waited-for events. 

\subsection{SMT-Like Model Analysis}
One of the advantages of scenario-based modeling is its amenability to
formal verification with appropriate tools, both by exhaustive model
checking traversing all paths, and by using domain knowledge for
Satisfiability Modulo Theory (SMT)
verification~\cite{HaLaMaWe11,HaKaKaMaMiWe13}. This is accomplished by
virtue of the abstraction and encapsulation of domain-specific
processes, actions and conditions as events and states.  System
complexity thus emerges from the composition of relatively small intuitive
scenarios reflecting individual requirements, and not from the
intricate conditional flow of delicate and sensitive processes with
numerous steps.

Our experiments have shown that ChatGPT is able to leverage this kind
of abstraction and encapsulation to identify cases that a
specification either omitted or handles incorrectly. For example, we
presented ChatGPT with the following three requirements for an
autonomous vehicle: (i) always stop at a red light; (ii) always obey
a police person's instructions; (iii) never injure a person. ChatGPT
readily recognized that these requirements may be in conflict with
each other. Given the safety property of not entering an intersection
when the traffic light is red, it pointed out that the AV may be unable to
stop if the road is icy, that it may enter the intersection
involuntarily if after stopping it is hit from behind by a car that
did not stop, and, furthermore, that it may injure a person without
moving, if a person walks behind the AV, and another vehicle hits the
person who is then thrust against the AV.

In a mathematical, SMT-like analysis, ChatGPT was able to identify a 
particularly interesting execution path. 
We presented it with a system consisting of the following
four scenarios: the first requests flashing  a red light briefly in
response to an external time tick event, and then waits for any event; the
second does the same with a green light; the third scenario blocks the
event of flashing the red light every third tick, and another blocks
flashing the green light every five ticks.  We then added the
requirement that at least one light must be
flashed between any two consecutive time ticks. 
ChatGPT was able to respond that
the requirement is not satisfied, and that every 15 time ticks
($3 \cdot 5$) both events will be blocked.  ChatGPT apparently recognized this
by arithmetic calculation and not by step-by-step simulation, as it
was able to quickly reach the same conclusion for arbitrarily large
numbers.

Here is an excerpt from this conversation, after describing the
scenarios. When prompted with

\begin{mdframed}[backgroundcolor=blue!10]
  now a new requirement came in: always, after every tick, \emph{at
    least one light} must be flashed. Does the combined system of red
  flashing and green flashing with the 3 and 5 restrictions satisfy
  the requirement?
\end{mdframed}

ChatGPT replies:

\begin{mdframed}[backgroundcolor=purple!10]
  To analyze whether the combined system satisfies the new requirement
  that at least one light must flash after every tick, we must
  consider the synchronization points of both the red and green
  systems with their respective restrictions:

  The red system flashes on every tick, except every third one.  The
  green system flashes on every tick, except every fifth one.  If we
  consider the sequence of ticks and apply both restrictions, the
  combined system would not flash a light on a tick that is both a
  multiple of three and five --- specifically, on ticks that are
  multiples of fifteen, since 15 is the least common multiple of 3 and
  5. At such ticks, both systems would be blocking their respective
  lights from flashing due to their individual restrictions.

  Therefore, on ticks that are multiples of fifteen (like the 15th,
  30th, 45th tick, and so on), neither light would flash, violating
  the new requirement that at least one light must flash after every
  tick.

  In conclusion, the combined system of red flashing and green
  flashing with the restrictions of not flashing on ticks divisible by
  three and five, respectively, does not satisfy the requirement that
  at least one light must be flashed after every tick.
\end{mdframed}

ChatGPT  also successfully discussed generalizations of this computation, including 
calculating  the number of composite system states that would have to be traversed in exhaustive model checking. 

It is interesting to note though, that while the LLM was able to apply
its logic to find system execution paths with special properties, it
did not do well when asked to list \emph{all} the execution paths
possessing a certain property. For example, we presented it with a
4-by-4 grid of cells, and asked for the steps that a robot may take
from (0,0) at the bottom left to (3,3) at the top right, using the
events for right, left, up, and down motions. We also required that
the path remains within the grid. ChatGPT was able to list paths like
\emph{right,right,right,up,up,up} and
\emph{right,up,right,up,right,up}, but when asked to list all the
paths that go through (1,3) it failed, listing several paths that
do not reach the desired target,
including \emph{up,up,up,right,down,down,right,up,right}. ChatGPT is thus not yet ready to
carry out exhaustive model checking on scenario-based systems.  This
is not really an  issue, however, since with the proposed methodology 
verification of the LLM's outputs will be carried out
independently of the LLM environment.

\subsection{Identifying Under-Specification}
Even when the created model addresses  the stated
requirements perfectly, as may be possible with synthetic or reduced problem
descriptions, there may be additional areas of the broader problem
domain that can or must be handled. In response to appropriate prompts
ChatGPT was able to significantly enrich a developed model, or helped
restate the explicit boundaries and expected scope of the intended
system. 

For example, in~\cite{HaAsFoLiMaSz23}
we presented ChatGPT with a
scenario-based model that was developed for science teaching
purposes. The model describes how bees depend on the ambient temperature for
switching between hibernation and being active, flower plants depend
on hours of daylight in order to start blooming, and active bees depend on
blooming flowers for survival. In the original model, average
temperatures and number of daylight hours were parameters that
students could adjust to see how certain combinations of settings
might be sustaining while others can cause the bees to die out due to
hunger. 

When we asked ChatGPT to offer enhancements to the education
model, it suggested to consider weather-related variations that modify
the pre-specified fixed average temperature and the number of daylight
hours for a given time of year.

\subsection{Explainability}
Given a certain observed or expected behavior of the system, ChatGPT
is able to produce a natural language description of the cause of that
 behavior, finding a reasonable balance between the arcane
details of the complete execution trace, and the overall requirement
that "this is what the system was supposed to do".

For example, following a paragraph on the semantics of SB modeling  
we presented the following prompt:

\begin{mdframed}[backgroundcolor=blue!10]
  In a system with two scenarios, one waits for time tick events and
  after each time tick requests a short beep (100 msec). The other
  scenario waits for time ticks, and after each such event requests a
  long beep (500 msec).  Each of these scenarios waits for both the
  short and long beep events, so that after each time tick, exactly
  one beep occurs.  Actuator scenarios actually cause the respective
  beeping action. Time tick events are injected by the environment
  every one second. A user reported that it watched the system for 20
  seconds, and never heard a long beep. Can you explain why this may
  have happened?
\end{mdframed}

ChatGPT response was:

\begin{mdframed}[backgroundcolor=purple!10]
  The user's report of not hearing a long beep in a system with two
  scenarios---one requesting a short beep after each time tick, and
  the other requesting a long beep after each time tick---can be
  explained through the principles of Scenario-based Modeling
  (SBM).

  In SBM, when multiple scenarios request different events
  after a shared trigger event, the execution environment selects one
  of these requested events to occur. The selection can be arbitrary
  or follow a specific strategy, such as alphabetic or lexical order,
  priority, random, round-robin, etc. In this case, if the event
  selection strategy prioritizes the short beep over the long beep
  (for example, due to alphabetic order, where ``short beep'' comes
  before ``long beep''), then the short beep will always be selected
  to occur after each time tick, and the long beep will never be
  executed.  Since time tick events are injected by the environment
  every second, and both scenarios are waiting for this event, each
  time tick leads to both scenarios being activated. However, if the
  system consistently selects the short beep event over the long beep
  event after each time tick, the scenario requesting the long beep
  will never progress to the state where it actually triggers the long
  beep. As a result, only short beeps will be heard.

  In summary, the
  user's observation of not hearing any long beeps can be attributed
  to the event selection mechanism within the SBM framework, which in
  this case consistently favors the short beep event over the long
  beep event. This could be due to the selection strategy used by the
  system (like alphabetic order) or other operational logic specifics.
\end{mdframed}

We then challenged ChatGPT again:

\begin{mdframed}[backgroundcolor=blue!10]
  Thank you. Assume that the developers changed the selection
  algorithm and they now select events randomly. The user tested the
  system many times, for hours, and never heard a long beep. Can you
  try to explain?
\end{mdframed}

To this, ChatGPT produced a very comprehensive answer,
covering possible issues with application scenarios, the execution infrastructure, and human-machine interaction.  Here are some excerpts:

\begin{mdframed}[backgroundcolor=purple!10]
  If the event selection algorithm was changed to randomly select
  between the short beep and long beep events, but the user still
  never hears a long beep, there are several potential explanations:

  Implementation Error in Random Selection: The most straightforward
  explanation could be a bug or implementation error in the random
  selection algorithm. It might not be truly random, or it could be
  incorrectly favoring the short beep event due to a coding mistake.

  Issue with the Long Beep Scenario: There might be a problem with the
  scenario that requests the long beep. This could be a coding error
  or a logical flaw in the scenario itself, preventing it from
  properly requesting the long beep event.




  User Perception or Reporting Error: Finally, there could be an issue
  with how the user is perceiving or reporting the beeps. They might
  be missing the long beep due to distraction, ambient noise, or
  misunderstanding what constitutes a long beep in the context of
  this system.

To diagnose the issue accurately, developers would need to conduct
thorough testing and debugging, looking into each of these potential
areas to find where the problem lies.
\end{mdframed}

While such explanations are impressive, they were not perfect. We had cases where we presented ChatGPT with well specified scenarios and a description of an observed  execution log, i.e., an event sequence, but ChatGPT was not able to retrospectively produce a step by step play-out 
sequence that would generate this execution log. Instead, it provided incorrect explanations for that event sequence. 
While frequently providing  ChatGPT with better  reminders and interactive practice of its  knowledge of semantics and play-out
may remedy such issues, this caveat is a reminder that all outputs produced by the LLM must be formally checked. 

\subsection{Accommodating Semantic Flexibility}
Most software development studios are tied to specific languages and
their associated semantics. In our experiments, ChatGPT was able to
accommodate, and discuss, alternative semantics.  

For example, in the
water tap example in Section~\ref{sec:background}, when the
scenario \textsc{AddHotWater} is in any of the states where it requests
\textsc{AddHot}, it cannot react to \textsc{WaterLow}, since it is not waiting
for that event. By contrast, in the semantics of the LSC
scenario-based language, the infrastructure constantly listens for all
events that are waited for in the first state of all scenarios. When
such an event occurs, the infrastructure instantiates another copy of
the scenario. In fact, from our first textual descriptions of SBP,
ChatGPT understood this semantics to be the default.  

In another
example, we  asked ChatGPT to generate scenarios for Quicksort.
Before starting, it commented that it will be hard, as classical
solutions are recursive.
We then pointed out to ChatGPT that there is a published implementation that is
iterative, not recursive~\cite{HaMaYe21}, that is 
structured as instructions to human workers arranging cars in an
automobile dealership parking lot according to, say, 
window-sticker price, where
each employee had one narrow role. ChatGPT readily accommodated the
new mindset and produced the desired scenario-based specification.

\subsection{Interactive Mutual Learning}
In our experiments, we noticed that ChatGPT learns from multiple
prompts, discussions and exploration better than from concise or
detailed descriptions. We believe that the same may hold for software
and system developers. Interactive, agile development processes may
not be just trial and error, or spiral convergence to and discovery of
a predefined but poorly specified goal. Rather, they are often constructive
processes, where stakeholders and developers build their wishes and
plans, as they refine their own understanding of the environment, the
systems, their needs, and their future interactions with the
system. 

An important part of this refinement is producing more
explicit definitions of elements that are outside  the scope of the
system. In contrast, such definitions are often totally absent from classical
system specifications.

\section{\uppercase{Related Work}}
\label{sec:relatedWork}

Recent advances in LLM-based chatbots have made a considerable impact
on numerous domains. Researchers and engineers are
now examining the potential applications of this technology in
education~\cite{AlDiJoLo23}, music~\cite{LuXuKaYuXiTaBi23},
academia and libraries~\cite{LuWa23}, healthcare~\cite{LiDaKlEg23}, and many
other areas. 

Within the field of software engineering, which is our subject
matter here, attempts have been made to apply chatbots to evaluate the
quality of code~\cite{BuOzAyTu23}, to correct bugs~\cite{SuSh23}, and
to generate code automatically or
semi-automatically~\cite{FeVaChZhQiCh23,DoJiJiLi23}. The general consensus appears to
be that chatbots will play a key role in code generation in years to
come. Our work here outlines a possible path towards allowing this
integration in a safe and controlled manner.

Our proposed methodology for integrating ChatGPT into the software
engineering process leverages the large body of existing work on
scenario-based modeling~\cite{HaMaWe12,DaHa01}. Specifically, we
propose to make use of the amenability of SBM to formal analysis
techniques~\cite{HaKaLaMaWe15,HaKaMaWe15}, such as
verification~\cite{HaLaMaWe11,HaKaKaMaMiWe13}, automatic
repair~\cite{HaKaMaWe12}, and synthesis~\cite{GrGrKaMa16}. Despite our
focus on SBM, other modeling approaches, with similar traits, could be
used in a similar manner.

Finally, our work here can be regarded as another step towards the
\emph{Wise Computing} vision~\cite{HaKaMaMa16,HaKaMaMa16b,HaKaMaMa18},
which seeks to transform the computer into an active member of the
software engineering team --- raising questions, making suggestions
and observations, and carrying out verification-like processes, even without explicitly being asked to do so.

\section{\uppercase{Conclusion}}
\label{sec:conclusion}

The appearance of large language models, and the subsequent release of advanced
chatbots, is a major development, and it is likely to revolutionize
the domain of software engineering in coming years. However, because
of inaccuracies and errors that are inherent to the outputs of these
chatbots, such an integration must be performed with care. We outline
here a possible method for such an integration, which makes use of the
advanced features of chatbots, but which also puts an emphasis on
inspecting and analyzing the results of the integration. We are
hopeful that our work  will trigger additional research in this
important direction.

Moving forward, we plan
to continue this work along several axes. First
and foremost, we intend to implement the necessary tools and
environments needed to fully integrate ChatGPT with SBM, and then
use these tools and environments in large, real-world case studies
that will demonstrate the usefulness of the approach as a whole. 

In addition, we expect that this line of work will require us to enhance
and modify existing tools, both on the SBM said and on the chatbot
one. For instance, with the current version of ChatGPT, every
conversation starts from a blank slate, whereas for the ongoing
development of a system, as part of the Wise Computing vision, it would
be more useful to have the chatbot remember and use previous
conversations. This could be achieved, for instance, by summarizing
each conversation as it ends, and then feeding these summaries back to
the chatbot when a new conversation starts. 
In fact, with the newly announced GPTs feature introduced in ChatGPT one can build a chatbot that is customized specifically for developing SB models and programs.

Ideally, LLMs will be able learn immediately from ongoing conversations, yet they will do so selectively, learning over time, to select what should be retained in each conversation and for how long.

These developments can also be beneficial in a broader perspective: prompt engineering methods and practices that would be developed along the way for such interactive, incremental development may prove useful not only in teaching computers, but in enhancing the training and everyday communications of human engineers.

\section*{\uppercase{Acknowledgements}}

The work of Harel, Marron and Szekely was funded in part by an NSFC-ISF grant to DH,  issued jointly by the  National Natural Science Foundation of China (NSFC) and the Israel Science Foundation (ISF grant 3698/21). Additional support was provided by a research grant to DH from the Estate of Harry Levine, the Estate of Avraham Rothstein, Brenda Gruss, and Daniel Hirsch, the One8 Foundation, Rina Mayer, Maurice Levy, and the Estate of Bernice Bernath.

The work of Katz was partially funded by the European Union (ERC,
VeriDeL, 101112713). Views and opinions expressed are however those of
the author(s) only and do not necessarily reflect those of the
European Union or the European Research Council Executive
Agency. Neither the European Union nor the granting authority can be
held responsible for them.

\bibliographystyle{apalike}
{\small
\bibliography{sbp_and_gpt}}

\end{document}